# Peano count trees (P-trees) and rule association mining for gene expression profiling of DNA microarray data

Valdivia Granda, W.A [1]*, Perrizo, W.[2], Larson, F[3] and Deckard, E.L.[1]

[1] Department of Plant Sciences, Loftsgard Hall Office 460-C, North Dakota State University, Fargo, ND 58105, United States of America

[2] Department of Computer Science, Loftsgard Hall Office 460-C, North Dakota State University, Fargo, ND 58105, United States of America

[3] Information Technology Services, Loftsgard Hall Office 460-C, North Dakota State University, Fargo, ND 58105, United States of America

* Corresponding author: Willy.Valdivia@ndsu.nodak.edu

**Abstract**

The greatest challenge in maximizing the use gene expression data is to develop new computational tools capable of interconnecting and interpreting results from different organisms and experimental settings. We propose an integrative and comprehensive approach involving a "super chip" containing data from microarray experiments performed on different species subjected to hypoxic and anoxic stress. A patented data mining technology called Peano count tree (P-tree) is used to represent genomic data in multi-dimensions. Each spot of a microarray is presented as a pixel with its corresponding red and green feature bands. Each band is stored separately in a reorganized 8-separated (bSQ) file. Each bSQ is converted to a quadrant base tree structure (P-tree) from which the "super-chip" is represented as expression P-trees (EP-trees) and repression P-trees (RP-trees). The use of Association Rule Mining is proposed as a mean to derive meaningful rules of gene interaction and to organize signal transduction pathways taking in consideration evolutionary aspects. We argue that the genetic constitution of an organism ($K$) is represented by the total number of genes belonging to two different groups. The group $X$ constitutes genes ($X_1,...,X_n$) and they can be represented as 1 or 0 depending on whether the gene was expressed or not. The function of many of these genes is conserved among organisms. The second group of $Y$ genes ($Y_1,...,Y_n$) are expressed or repressed at different levels. These genes have a "very high expression", "high expression", "very high repression" or "high repression" levels. However, many genes of the group $Y$ are specie specific and modulated by the products and combination of genes of the group $X$. In this paper we introduce the bSQ and P-tree technology; the biological implications of association rule mining using $X$ and $Y$ gene groups and some current advances in the integration of this information using the BRAIN architecture.

**Biological implications of hypoxic and anoxic stress**

Adequate supply of oxygen is essential to all higher organisms for an adequate provision of energy and survival. Oxygen serves as the terminal electron acceptor in the mitochondria oxidative phosphorylation and it is a substrate in several enzymatic processes. However, cellular oxygen concentration varies during ontogenesis and it is dependent on different environmental conditions. How cells sense, tolerate and adapt to hypoxic and anoxic stress is a central question in biology. Despite the importance of this process in cell development, the information available remains incomplete. Recent progress in understanding the molecular bases of the effects of reduced oxygen concentration in humans, animals and plants has demonstrated common sensing and tolerance pathways. We argue that although the information for one organism regarding the hypoxic stress may be limited, integrating the data from different experiments performed on several species can provide a great deal of information and may lead to new discoveries.

Genomic research using DNA microarray has become a cross-disciplinary endeavor in which the one gene-at-a-time is replaced by the global analysis of thousands of genes. The rapid progress in microarray technologies, the current data standardization efforts, and the increasing accessibility to microarray chips and equipment, has motivated a growing consensus for the need of public repositories of both microarray images and data. However microarray technology raises new computational and statistical challenges including: 1) the limited capability of current tools to interrogate a large microarray database or multiple databases generated from different experiments and different species, 2) semantic interoperability of different genomic data systems where each system (or object of a system) can map its own conceptual model to the conceptual model of other systems, 3) data transfer and data compression issues and 4) the development of new algorithms for a better interpretation of the biological processes.

*The bSQ Format and the P-tree Data Structure for Microarray Data*
The expression level of each gene is indirectly recorded by the measurement of the fluorescence level emitted by each dye (red/green) attached to the cDNA. Each spot on the microarray emitting a signal is a pixel with byte number ranging from 0 to 255. Different bits can make different contributions to the values that are used for gene expression profiling. Therefore, a microarray image can be organized into an 8-separated bit sequential (bSQ) format. The intensity of each band (red/green) is stored in two separated bSQ files. The primary key attribute of the bSQ format consists of the pixel location (x-y coordinates of the spot on the microarray) and its corresponding gene identification. The subsequent attributes consist of the bSQ values for each signal. There are several reasons to use the bSQ format. First, different bits have different degrees of contribution to the intensity value. In some applications, we do not need all the bits because the high order bits give us enough information. Second, the bSQ format facilitates the representation of a precision hierarchy. Third, the bSQ format facilitates better data compression. This point is relevant for the integration of genomic data and the performing of faster data mining applications. Fourth, and most importantly, the bSQ format facilitates the creation of an efficient, rich data structure denominated Peano Count Tree (P-tree) that accommodates algorithm pruning based on a one-bit-at-a-time approach. The Peano Count Tree is a lossless tree representation where the root of a P-tree contains the 1-bit count of the entire bit-band representing the microarray spot. At the next level, each quadrant is partitioned into sub-quadrants and their 1-bit counts in raster order constitute the children of the quadrant node. This construction is continued recursively down each tree path until the sub-quadrant is pure (entirely 1-bits or entirely 0-bits), which may or may not be at the leaf level (1-by-1 sub-quadrant). Our approach is to recursively divide the entire image into quadrants and then record the count of 1-bits in each quadrant, thus, forming a quadrant count tree (Fig. 1).

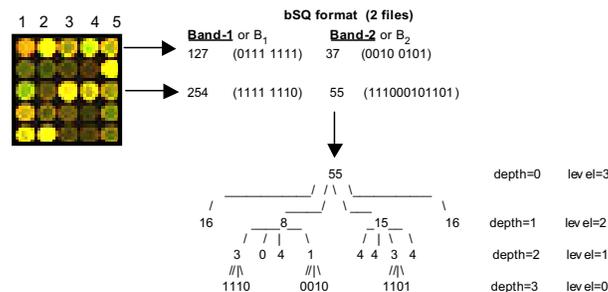

Fig. 1. Representation of gene expression levels in a bSQ format

We give a very simple illustrative example with only 2 data bands for a microarray image having only 2 rows and 2 columns (both decimal and binary representation are shown) (Fig. 2).

| BAND-1 | | BAND2 | | BSQ format (2 files) |
|---|---|---|---|---|
| 254 *(1111 1110)* | 127 *(0111 1111)* | 37 *(0010 0101)* | 240 *(1111 0000)* | Band 1: 254 127 14 193 |
| 14 *(0000 1110)* | 193 *(1100 0001)* | 200 *(1100 1000)* | 19 *(0001 0011)* | Band 2: 37 240 200 19 |

| bSQ format (16 files) | | | | | | | | | | | | | | | |
|---|---|---|---|---|---|---|---|---|---|---|---|---|---|---|---|
| B11 | B12 | B13 | B14 | B15 | B16 | B17 | B18 | B21 | B22 | B23 | B24 | B25 | B26 | B27 | B28 |
| 1 | 1 | 1 | 1 | 1 | 1 | 1 | 0 | 0 | 0 | 1 | 0 | 0 | 1 | 0 | 1 |
| 0 | 1 | 1 | 1 | 1 | 1 | 1 | 1 | 1 | 1 | 1 | 1 | 0 | 0 | 0 | 0 |
| 0 | 0 | 0 | 0 | 1 | 1 | 1 | 0 | 1 | 1 | 0 | 0 | 1 | 0 | 0 | 0 |
| 1 | 1 | 0 | 0 | 0 | 0 | 0 | 1 | 0 | 0 | 0 | 1 | 0 | 0 | 1 | 1 |

Fig. 2. Two bands of a 2-row-2-column image and its bSQ formats

In the example of Fig. 1, the root level is labeled as *level 0*. The numbers at the next two levels (level 2) are, 16, 8, 15 and 16, are the 1-bit counts for the four major quadrants. Since the first and last quadrants are composed entirely of 1-bits (called a "pure 1 quadrant"), we do not need sub-trees for these two quadrants, so these branches terminate. Similarly, quadrants composed entirely of 0-bits are called "pure 0 quadrants" which also terminate. This pattern is continued recursively using the Peano or Z-ordering of the four sub-quadrants at each new level. Every branch terminates eventually (at the "leaf" level, each quadrant is a pure quadrant). If we were to expand all sub-trees, including those for pure quadrants, then the leaf sequence is just the Peano-ordering (or, Z-ordering) of the original raster image. Thus, we use the name Peano Count Tree (P-tree). This structure provides compression and embedded information that is needed to do genomic data mining. P-trees defined above can be combined using simple logical operations [AND, NOT, OR, COMPLEMENT] to produce additional P-trees from the original values from each band, *b*, and value, *v*, where *v* can be expressed in 1-bit, 2-bit,.., or 8-bit precision (Perrizo et al. 2001; Perrizo et al. 2001a). Using this approach we derive expression P-trees (EP-trees) and repression P-trees (RP-trees) defined by the red/green intensity of each spot that are significantly above or below of the reference genes spotted on a microarray.

P-trees are a lossless and compressed data structures that can be use to construct a "super chip" which is derived from multiple experiments (in our case data generated from researchers of Virtual Center for Hypoxic and Anoxic Research www.ndsu.edu/virtual-genomics). The Biologic Research Application and Information Network (BRAIN) integrates different sources of gene expression using a distributed data system (DDS). The DDS is a Java system that makes use of JDBC and XML to connect disparate databases and allow them to be queried as one system. DDS allows the integration of different data sources and the translation of different data formats such as CSV (comma separated variable) file, XML or the tables of relational databases. Through the use of drag and drop the user is able to associate input from various sources with existing tables and then generates the SQL needed to insert the data. While the systems allow each researcher keep and develop their own database, it does not require changing their existing database schemas. Thanks to BRAIN gene expression data is represented in multidimensional fashion.

***Considerations to use association rule mining techniques for microarray data***
Association-rule mining is a widely used technique for large-scale data mining with applications in different areas including market basket research, insurance fraud investigation, climate prediction and remote sensing research.

An association rule is a relationship (X ⇒ Y) where X is the antecedent item set and Y is the consequent item set. In mining for these rules the user defines a threshold, called *confidence*, which the implication, measured in terms of a *conditional probability*, must exceed. The primary task here is to identify frequent item sets, that are sets of data points occurring with at least a minimum frequency, called *minimum support*. Once the frequent item sets are identified, the association rules are formulated. To accomplish this task different algorithms have been developed including Apriori, Charm, FT-growth, Closet, MagnumOpus, etc. It is argued that today's association rule mining algorithms are able to efficiently prune data sets (Hipp et al. 2000). However, one of the greatest challenges in use association rule mining for gene expression data analysis is the immense large theoretical rules that need to be considered. This is a consequence of all possible combinations of gene expression under different conditions (temporal, spatial and experimental). In addition, not all association rules discovered within a transaction set are interesting or useful (Zheng et al. 2001).

Microarray technology is still under development. However this technique provides us with three basic kinds of information: i) the gene expression or repression ii), gene expression or repression level and iii) the interaction of a particular gene with other genes. If we use integrative data architectures (like BRAIN) we can use association rule mining to determine the *minimum support* value for a gene and subsequently for a group of genes by counting their *occurrences* in a genomic database (previously defined as the "super chip"). Although here we do not present the performance of association rule mining algorithms in gene expression data mining, we propose the bases for its application. Association rule mining can be used for DNA microarray data mining in a hierarchical fashion if we take in consideration the complexity of biological systems. We argue that the genetic constitution of an organism ($K$) is represented by a total number of genes belonging to two different groups. There are constitutive genes expressed or repressed during the life span of an organism. These genes have a similar level of temporal and spatial expression and in many cases are conserved and have similar function in other species. They are the first gene group of our model (genes $X$) constituted by genes $X_1,...,X_n$ and represented as 1 or 0 depending if the gene was expressed or not. The other group in our model (genes $Y$) are expressed or repressed in spatial and temporal fashion (specific tissue, organ, developmental stage and environmental condition). The genes belonging to this group are specie and even genotype specific. This group of genes is represented in levels as very high expression, high expression, very high repression or high repression. However, $Y$ genes are modulated by the presence and combination of genes products of the group $X$.

**Remarks**

DNA microarray technology is used by different laboratories with the aim to measure the temporal and spatial gene expression of different organisms subjected to diverse experimental settings. As the data grows, the need for public repositories and integrative architectures is becoming evident. In this paper we presented the bSQ format and P-tree technology to generate lossless and "data mining ready" representations of microarray images. We take in consideration the complexity of biological systems and assume that genes can be separated in two different main groups and that association rule mining can be applied for gene expression data analysis. However the bSQ and P-tree technology can be used with other classification or clustering techniques.

**References:**
1. Hipp, J., Güntze, U., Makheeizade, G. 2000. Algorithms for association rule mining – A general survey and comparison. SIGKDD (2 )1:58-64.
2. Perrizo, W., Ding, Q., Ding, D., Roy, A. 2001. On mining satellite and other Remetely Sensed Images. DMKD. 33-44.
3. Perrizo, W. 2001. Peano Cunt Tree Technolgy. Technical Report NDSU-CSOR-TR-01-1.
4. Zheng, Z., Kohavi, R., Mason, L. 2001. Real World Performance of Association Rule Mining. KDD-2001.